\documentclass[10pt, a4paper, oneside, reqno]{amsart}

\makeatletter
\def\@settitle{\begin{center}%
		\baselineskip14\p@\relax
		\normalfont\LARGE\scshape\bfseries
		\@title
	\end{center}%
}
\makeatother

\makeatletter

\def\subsection{\@startsection{subsection}{2}%
	\z@{.5\linespacing\@plus.7\linespacing}{.5\linespacing}%
	{\normalfont\bfseries}}

\def\subsubsection{\@startsection{subsubsection}{3}%
	\z@{.5\linespacing\@plus.7\linespacing}{.5\linespacing}%
	{\normalfont\itshape}}

\usepackage[usenames, dvipsnames]{color}
\definecolor{darkblue}{rgb}{0.0, 0.0, 0.45}

\usepackage[colorlinks	= true,
			raiselinks	= true,
			linkcolor	= darkblue, 
			citecolor	= Mahogany,
			urlcolor	= ForestGreen,
			pdfauthor	= {},
			pdftitle	= {},
			pdfkeywords	= {},
			pdfsubject	= {},
			plainpages	= false]{hyperref}

\usepackage{dsfont,amssymb,amsmath,graphicx} 
\usepackage{amsfonts,dsfont,mathtools,mathrsfs,amsthm} 
\mathtoolsset{showonlyrefs=true}
\usepackage[amssymb, thickqspace]{SIunits}
\usepackage{fancyhdr,setspace}

\usepackage[numbers,compress]{natbib}
\usepackage{nicefrac}       
\usepackage{microtype}      
\usepackage{tabto}

\usepackage{enumerate,enumitem}
 \usepackage{caption}

\usepackage{bbm}

\allowdisplaybreaks
\date{\today}

\newcommand{\T}{t\in\mathcal{T}}
\newcommand{\TT}{t\in\mathcal{T}_{H}}
\newcommand{\N}{n\in\mathcal{N}}
\newcommand{\R}{\mathbb{R}}

\newcommand{\Z}{Z_{n}(t)}
\newcommand{\ZK}{Z^{K}_{n}(t)}
\newcommand{\q}{q_{n}(t)}
\newcommand{\Y}{Y_{n}(t)}
\newcommand{\Q}{Q_{n}(t)}
\newcommand{\XP}{x_{n}(t+1)}
\newcommand{\X}{x_{n}(t)}

\graphicspath{{images/}}

\usepackage{url}            

\usepackage{algorithm}
\usepackage{algorithmic}

 \usepackage[foot]{amsaddr}

\usepackage{subfig}

\usepackage{algorithmic}
{

{
		\theoremstyle{plain}
		
	}
	{
		\theoremstyle{plain}
		
	}
	{
	\theoremstyle{plain}
	
}
	{
		\theoremstyle{plain}
		
	}
	{
		\theoremstyle{plain}
		
	}
{
		\theoremstyle{plain}
		
	}
{
		\theoremstyle{plain}
		
	}

\title[]{Distributed joint dynamic maintenance and production scheduling in manufacturing systems:\\ 
Framework based on model predictive control and Benders decomposition}

\author[1]{Pegah Rokhforoz$^1$}
\author[3]{Olga Fink$^{2}$}
\address[1]{Chair of Intelligent Maintenance Systems, ETH Zurich, Switzerland, and School of Electrical and Computer Engineering, University of Tehran, Tehran, Iran.}
\address[2]{Chair of Intelligent Maintenance Systems, ETH Zurich, Switzerland. Corresponding author, email address:ofink@ethz.ch.}

\begin{document}

\maketitle

\begin{abstract}
Scheduling the maintenance  based on the condition, respectively the degradation level of the system leads to improved system's reliability while minimizing the maintenance cost. Since  the degradation level changes dynamically during the system's operation, we face a dynamic maintenance scheduling problem. In this paper, we address the dynamic maintenance scheduling of  manufacturing systems based on their degradation level. The manufacturing system consists of several units with a defined capacity and an individual dynamic degradation model, seeking to optimize their reward. The units sell their production capacity, while maintaining the systems based on the degradation state to prevent failures. The manufacturing units are jointly responsible for fulfilling the demand of the system. This induces a coupling constraint among the agents. Hence, we face a large-scale mixed-integer dynamic maintenance scheduling problem. In order to handle the dynamic model of the system and large-scale optimization, we propose a distributed algorithm using model predictive control (MPC) and Benders decomposition method. In the proposed algorithm, first, the master problem obtains the maintenance scheduling for all the agents, and then based on this data, the agents obtain their optimal production using the distributed MPC method which employs the dual decomposition approach to tackle the coupling constraints among the agents. The effectiveness of the proposed method is investigated on a case study.
\end{abstract}

\section{Introduction}
Optimal management of manufacturing systems has recently been addressed from many different perspectives. Several research studies focused on production planning in the presence of demand uncertainty \cite{Chen_2012}, \cite{Aouam_2018}, network transportation \cite{Amiri_2006designing}, \cite{feng2018two}, and maintenance scheduling \cite{he2017integrated}, \cite{Alimian_2020}. The performance of the manufacturing system depends on the quality of the production and also the maintenance strategy of the manufacturing machinery and equipment to prevent system's breakdowns and consequently production disruptions \cite{Sarkar_2011}. Moreover, the supply agreements should be satisfied in all operating time even if some units undergo maintenance. Hence, simultaneous maintenance and production scheduling play an important role in operating and managing of manufacturing systems which are considered in many research papers such as in \cite{Kang_2018}, \cite{Polotski_2019}.

Deterioration of industrial and manufacturing systems can significantly affect the production quality. Not maintaining critical degradation levels on time can result in severe production disruptions. Deterioration rate depends on many different factors, including material, system design, manufacturing and process parameters, load, stress and environmental factors \cite{Pahl_2014}, and \cite{Janssen_2016}. Several research studies addressed the problem of modeling deterioration rates in manufacturing systems, including \cite{Ahmadi_2011}, \cite{Rivera_2018}. In \cite{Ahmadi_2011} the deterioration rate is modeled as an non-decreasing function of damage process where the base-line rate depends on the age of the system.  \citet{Rivera_2018} propose an increasing function for modeling the deterioration rate which depends on several factors such as usage, corrosion and environment.

Recently, real-time monitoring of the degradation level of the assets has become more cost efficient and also more effective due to the technological advancements in condition monitoring technology. This enables early fault detection enabling maintenance to be performed before the critical failure occurs \cite{fink_2020}. Depending on the deterioration rate, also inspections in discrete time intervals may be sufficient to efficiently monitor the evolution of the system condition.  

Maintenance decision making is a critical action to improve the systems' reliability and availability during their useful lifetime. Moreover, since the system's degradation evolves dynamically, dynamic predictive maintenance scheduling based on the deterioration state of the system leads to improved system's performance compared to preventive maintenance. Preventive maintenance typically does not consider the system's condition and the potential dynamic changes in the condition \cite{Omshi_2020}. 

Several studies have addressed dynamic maintenance scheduling \cite{Alaswad_2017}. \citet{Makis_2015}, and \citet{Tang_2015} formulate the condition-based maintenance problem as a Markov decision model and obtain the maintenance scheduling using an optimization approach which minimizes the long-run expected average cost. Dynamic maintenance scheduling using dynamic programming is proposed in \cite{Ghasemi_2007}, \cite{Ghasemi_2008}, where the deterioration state is formulated using hidden Markov decision model with some uncertain parameters. Furthermore, the dynamic predictive maintenance policy for multi-component systems is developed in \cite{Van_2013}. The policy minimizes the maintenance cost per unit for a long horizon time. The dynamic maintenance scheduling is then updated based on the new information on the degradation state and the remaining useful life (RUL) time of the system. \citet{wang2016} propose a multi-state maintenance scheduling using model predictive control (MPC) where the control input is obtained by predicting the future state based on the current information of the system. There are a few papers that address dynamic maintenance scheduling in the manufacturing system.  \citet{kroning2013dynamic} propose a model-based approach which obtains the dynamic maintenance scheduling for a supply chain system for a short time horizon. \citet{boukas2001production} propose maintenance and production scheduling for one unit whose system state is modeled with a continuous-time Markov decision process. The production and maintenance rates are optimized based on the Markov decision process. The preventive dynamic maintenance scheduling and production control policy is proposed in \cite{Kang_2018} such that the minimum total production cost is achieved. 

The research studies described above obtain the maintenance scheduling by solving large-scale centralized optimization problems. This approach has a high computational cost. Furthermore, the central system needs to have access to all the unit's information, including their system states, deterioration functions and individual disruption and maintenance costs. 

There have been some research studies that propose distributed dynamic maintenance scheduling where the agents (sub-components) decide about their optimal maintenance individually, such as \cite{Aissani_2009}, \cite{yousefi_2020}, and \cite{Paraschos_2020}. \citet{Aissani_2009} propose a multi-agent maintenance scheduling in a petroleum system using reinforcement learning (RL). RL is also used in \cite{yousefi_2020} to obtain maintenance decisions for sub-components. The authors consider finite discrete values for the degradation state and solve the problem using Q-learning.  The authors of \cite{Paraschos_2020} propose a dynamic maintenance and production scheduling using a RL approach by considering one agent with discrete action space. These studies do not consider coupling constraints among the agents that ensure that the total demand is fulfilled. Furthermore, the approaches that consider continuous state-space dynamics are computationally expensive. 

In this paper, we address the problem of joint production and maintenance scheduling for a manufacturing system that comprises several production units. The central manufacturing system with its distributed production units is modeled as multi-agent system. In this multi-agent system, each agent has its individual deterioration dynamic and the agents have to collectively satisfy the contractual product delivery commitments of the manufacturing system. First, we model the deterioration state of the agents as a linear dynamic system using a Markov decision process. Second, we model the problem as a large-scale mixed-integer quadratic programming (MIQP) optimization problem. Third, we propose a framework based on a distributed algorithm using a combination of model predictive control (MPC) and Benders decomposition approach. The proposed approach tackles the large-scale problem using the dual decomposition method to relax the coupling constraint (demand constraint) among the agents. The MPC  predicts the control input (maintenance and production) based on the current information of the system for a given time period. It  applies the first input to the system, and obtains new updated information of the system. This procedure continues until the end of the decision time period. Hence, MPC obtains the control input online in real time. Thereby, it can handle the system uncertainty. This is one of the advantages of the proposed approach. We evaluate the performance of the proposed framework by comparing the optimality and computational cost of our proposed approach with the centralized MPC and the centralized method. 

The main contributions of the paper are the following:

1) We address the joint dynamic maintenance and production planning for multi-agents manufacturing systems.

2) We model the deterioration dynamic as a Markov decision model and formulate the problem as a mixed-integer quadratic programming optimization problem.

3) We propose the distributed algorithm to obtain the maintenance and production planning for large-scale manufacturing systems using the MPC and Benders decomposition approach which handle the large-scale optimization, dynamic model, and coupling constraint among the agents.

To the best of our knowledge, this is the first paper that develops a distributed algorithm which obtains the predictive maintenance and production planning for a long-term prediction horizon for a large-scale manufacturing system with dynamic agents.

The paper is organized as follows. We first describe a preliminary on MPC approach in Section \ref{sec:preliminary}. Then, the problem formulation is elaborated in Section \ref{sec:problem}. The proposed distributed framework is presented in Section \ref{sec:solution}. The simulation results are presented in Section \ref{sec:simulation}, and the concluding remarks are made in Section \ref{sec:conclusion}.

\textbf{Notation.} Given vectors $y(t)\in{\R^{n}}$, ${y(t)}^\top$ represents the transpose of $y(t)$. $\bold{y}=\bold{col}(y(1),...$ $,y(T))$ $=[{y(1)}^\top,\cdots, {y({T})}^\top]^\top$. Throughout the text, the index of iteration is denoted by superscript as $y^{k}$ and the power is denoted by parenthesis as $(y)^{k}$.

\section{Preliminary on Model predictive control}
\label{sec:preliminary}
Let $x(t)\in{\R^{s}}$ and $u(t)\in{\R^{c}}$ be the state and control input of the system at time $t$. Let us consider that the states of the system evolve through a linear dynamic as follows:

\begin{equation}
    x(t+1)=Ax(t)+Bu(t),
    \label{eq:dynamic}
\end{equation}
where, $A\in{\R^{s\times{s}}}$ and $B\in{\R^{s\times{{c}}}}$ are the system matrix and input matrix, respectively.  

 Let $J(x(t),u(t)):\R^{s}\times\R^c\rightarrow{\R}$ be the cost function at time $t$. The central system seeks to find the control input $\bold{u}=\textbf{col}(u(1),\cdots,u(T-1))$ and system states $\bold{x}=\textbf{col}(x(1),\cdots,x(T))$, which minimize the sum of the cost functions over the decision horizon time $\mathcal{T}={\{1,\cdots,T-1}\}$, as follows:
 
 \begin{equation}
 \begin{aligned}
     &\min\limits_{\bold{u},\bold{x}}\sum\limits_{\T}J(x(t),u(t)),\\
     &\text{S.b.}\quad \eqref{eq:dynamic},\quad F(x(t),u(t))\leq{0},\quad \T.
     \label{eq:mpc}
      \end{aligned}
 \end{equation}
We can obtain the control input and system's state at each decision instant $t$, using MPC approach in which one can solve the following receding horizon optimization approach over $\mathcal{T}_{H}={\{\tau,\cdots,\tau+H}\}$ by using the state of the system at time $\tau$:

\begin{equation}
 \begin{aligned}
     &\min\limits_{\bold{u},\bold{x}}\sum\limits_{\TT}J(x(t|\tau),u(t|\tau)),\\
     &\text{S.b.}\quad F(x(t|\tau),u(t|\tau))\leq{0},\quad \TT,\\
     &\qquad\quad x(t+1|\tau)=Ax(t|\tau)+Bu(t|\tau),\quad \TT,\\
     &\qquad\quad x(\tau)=x(\tau|\tau),
     \label{eq:mpc1}
      \end{aligned}
 \end{equation}
where $\bold{u}=\textbf{col}(u(\tau|\tau),\cdots,u(\tau+H|\tau))$, $\bold{x}=\textbf{col}(x(\tau|\tau),\cdots,x(\tau+H|\tau))$, and $H$ is the predicted horizon time. The variables $x(t|\tau)$ and $u(t|\tau)$ are the predicted state and control inputs at time $t$ based on the system's state information at time $\tau$, respectively. The receding horizon optimization \eqref{eq:mpc1} provides an open-loop optimal control sequence, and only the first control input $u(\tau|\tau)$ is applied to the system. Then, based on the new information of the system $x(\tau+1)$, by setting $\tau=\tau+1$, and using \eqref{eq:mpc1}, the next control input $u(\tau+1)$ can be obtained. This procedure continues until the control input at the end of the horizon $u(T-1)$ is obtained, \cite{Camponogara_2002}.

\section{Problem formulation}
\label{sec:problem}

In this section, we first model the deterioration dynamic of the units in the manufacturing system. We then model the objective function of the units or agents which determines the optimal joint maintenance and production scheduling for the long-term time horizon.

Let us assume that the manufacturing system consists of  $\mathcal{N}={\{1,\cdots,N}\}$ sub-systems which we denote as agents. We assume that the agents have a linear deterioration dynamic which is a function of their production and the performed maintenance as follows:

\begin{equation}
    \XP=(1-\Z)(A\X+B\q),
    \label{eq:dynamic deteroration}
\end{equation}
where $\X\in{\R}$, $\Z\in[0,1]$, and $\q\in{\R^{+}}$ are the deterioration state, the maintenance decision, and the production units, of agent $n$ at time $t$, respectively. Equation \eqref{eq:dynamic deteroration} implies that the deterioration state of the system is increasing when the agents are producing some products and the deterioration state returns to its initial value, which we consider as zero, after the  maintenance is performed.

The objective function of the agents is to maximize their reward function which is obtained by their production and minimize their production and deterioration costs. The aim of the proposed supply chain optimization problem is on the one hand to ensure that the deterioration states of the agents keep below defined threshold values by scheduling the maintenance before that threshold is reached, and on the other hand to make sure that the demand of the system is fulfilled while the agents perform maintenance at each decision time. Based on these boundary conditions, we can model the optimization problem as follows:

\begin{equation}
\begin{aligned}
    &\max\limits_{\bold{q},\bold{Z},\bold{x}}\sum\limits_{n\in\mathcal{N}}\sum\limits_{t\in\mathcal{T}}\big(\mathcal{P}({t})\q-a_{n}({\q})^2-b_{n}\q-\X^{T}\X\big)\\
    &\text{S.b.}\quad \text{C}_{1}:\sum\limits_{n\in\mathcal{N}}\q=D(t),\quad\T,\\
    &\quad\qquad \text{C}_{2}:\XP=(1-\Z)(A\X+B\q),\quad{\N},\quad\T,\\
    &\quad\qquad \text{C}_{3}:{(1-\Z){{q_{n}}^{\min}}\leq{\q}\leq{(1-\Z){{q_{n}}^{\max}}}},\quad{\N},\quad\T,\\
    &\quad\qquad \text{C}_{4}:\X\leq{G_{n}},\quad{\N},\quad\T,\\
        &\quad\qquad \text{C}_{5}:\Z\in[0,1],\quad{\N},\quad\T,\\
    \label{eq:main problem}
    \end{aligned}
\end{equation}
where $\bold{q}=\textbf{col}(q_{1},\cdots,q_{N})$, $\bold{q}_{n}=\textbf{col}(q_{n}(1),\cdots,q_{n}(T-1))$, $\bold{Z}=\textbf{col}(Z_{1},\cdots,Z_{N})$, $\bold{Z}_{n}=\textbf{col}(Z_{n}(1),\cdots,Z_{n}(T-1))$, $\bold{x}=\textbf{col}(x_{1},\cdots,x_{N})$,and $\bold{x}_{n}=\textbf{col}(x_{n}(1),\cdots,x_{n}(T))$, $\N$. $\mathcal{P}(t)$ is the price of the production at time $t$, $a_{n}$, and $b_{n}$ are the coefficients of the production cost of agent $n$. $D(t)$ is the load of the system at time $t$, $G_{n}$ is the threshold value for the deterioration state of agent $n$ which is determined based on the condition monitoring data of the system, $q_{n}^{\min}$ is the minimum production capacity, and $q_{n}^{\max}$ is the maximum production capacity. Constraint $\text{C}_{1}$ makes sure that the demand of the system is fulfilled. Constraint $\text{C}_{2}$ expresses the deterioration dynamic of the agents. Constraint $\text{C}_{3}$ ensures that the production volume is between its maximum and minimum capacities and it is zero during maintenance is performed. Constraint $\text{C}_{4}$ makes sure that the deterioration state is below its threshold value, and Constraint $\text{C}_{5}$ denotes that the maintenance decision is a binary variable.

Problem \eqref{eq:main problem} is a non-linear programming problem. We can convert it to the mixed-integer quadratic programming (MIQP) by introducing $\Y=\Z\X$, $\Q=\Z\q$, and using the big-$M$ method. We add the following constraints which make the reformulated problem equivalent to the non-linear term:

\begin{equation}
    \begin{aligned}
    &{0}\leq\Y\leq{\Z M},\\
    &\Y\geq{\X-(1-\Z)M},\\
    &\Y\leq{\X+(1-\Z)M},\\
     &{0}\leq\Q\leq{\Z M},\\
        &\Q\geq{\q-(1-\Z)M},\\
    &\Q\leq{\q+(1-\Z)M}.
    \label{eq:big m}
\end{aligned}
\end{equation}

Then, Equation \eqref{eq:main problem} is equivalent to the following MIQP:

\begin{equation}
\begin{aligned}
    &\max\limits_{\bold{q},\bold{Z},\bold{Y},\bold{Q},\bold{x}}\sum\limits_{n\in\mathcal{N}}\sum\limits_{t\in\mathcal{T}}\big(\mathcal{P}({t})\q-a_{n}({\q})^2-b_{n}\q-\X^{T}\X\big)\\
     &\text{S.b.}\quad\text{C}'_{1}:\XP=(A\X+B\q)-(A\Y+B\Q),\quad\N,\quad\T,\\
     &\qquad\quad \text{C}_{1}, \quad\text{C}_{3}-\text{C}_{5},\quad \eqref{eq:big m},\quad\N,\quad\T,\\
    \label{eq:main problem1}
    \end{aligned}
\end{equation}
where $\bold{Y}=\textbf{col}(Y_{1},\cdots,Y_{N})$, $\bold{Y}_{n}=\textbf{col}(Y_{n}(1),\cdots,Y_{n}(T-1))$, $\bold{Q}=\textbf{col}(Q_{1},\cdots,Q_{N})$, and $\bold{Q}_{n}=\textbf{col}(Q_{n}(1),\cdots,Q_{n}(T-1))$.

\section{Proposed distributed framework based on MPC and Benders decomposition methods}
\label{sec:solution}

In this section, we describe the  proposed distributed framework which is based on the MPC and Benders decomposition approach. Benders decomposition has been applied in several research studies to solve the mixed-integer programming problem in a distributed way, such as in \cite{Azad_2019}, \cite{cordeau_2000}.

\begin{figure}
    \centering
    \includegraphics[width=7in]{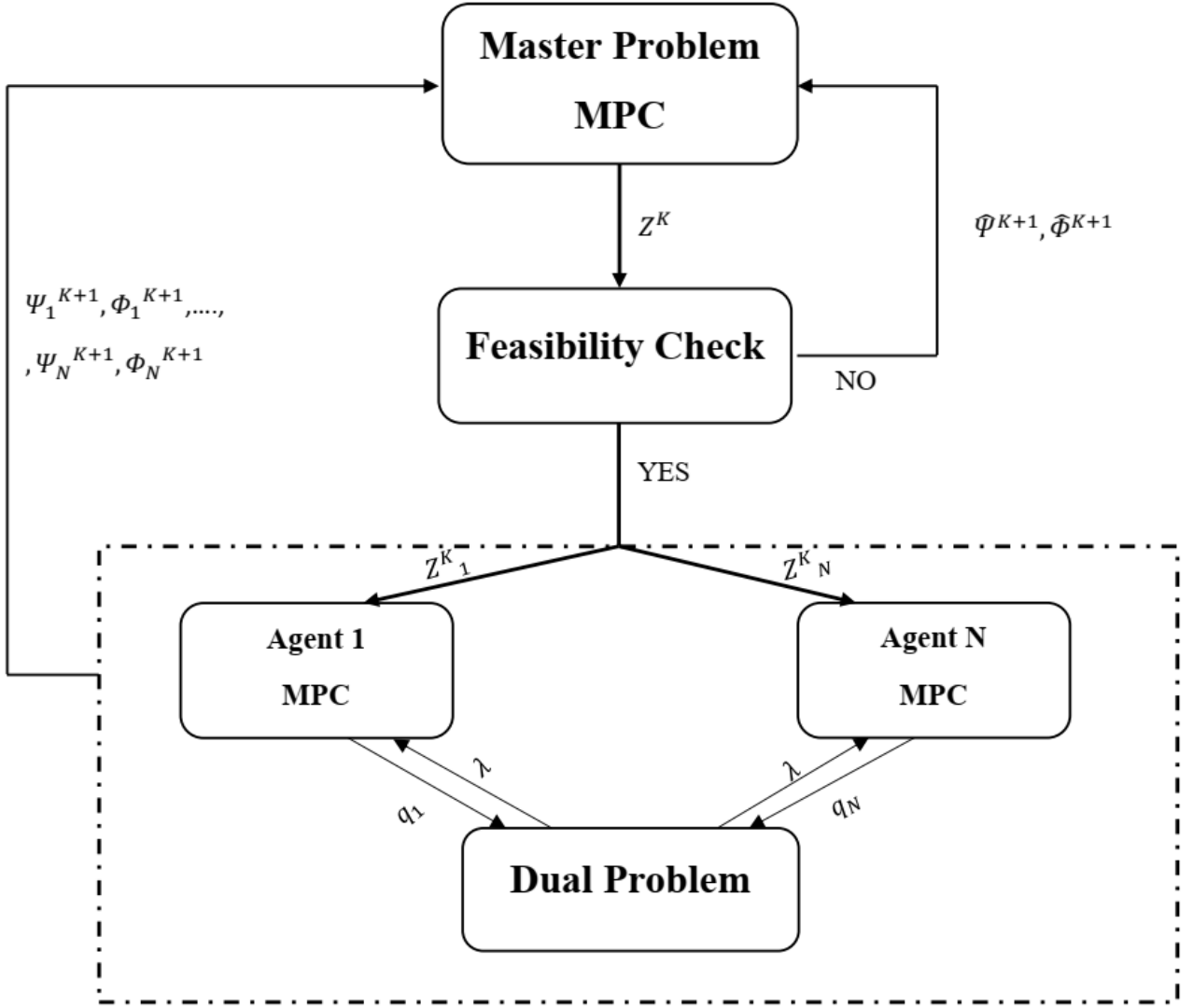}
    \caption{Schematic of the proposed iterative distributed framework.}
    \label{fig:master}
\end{figure}

The schematic of the proposed framework is illustrated in Figure \ref{fig:master}. The framework comprises three main parts: master problem, feasibility check, and agents' sub-problems. In this algorithm, first, the master problem obtains the maintenance decision and sends it to the agents. Then, the feasibility of the solution of the master problem is verified. If the solution is not feasible, some additional constraints which are known as feasibility cuts are added to the master problem. If the solution is feasible, the maintenance decisions are sent to the agents. In the agents' sub-problem part, each agent obtains its decision variables including the production units and the deterioration state using the distributed MPC approach which is obtained by applying the dual decomposition approach. Then, the optimality solution of the algorithm is checked and in the case that it does not satisfy the optimality conditions, some additional constraints known as optimality cut constraints are added to the master problem. This procedure continues until the optimality condition is satisfied.

\textbf{Master problem.} The maintenance decision can be obtained using the MPC approach by solving the following master problem:

\begin{equation}
\begin{aligned}
    &\min\limits_{\bold{Z}^{K},\bold{U}^{K}}\sum\limits_{\TT}U^{K}(t)\\
    &\text{S.b.}\quad\text{M}_{1}: [\phi^{k}_1(t),\cdots,\phi^{k}_{N}(t)]Z^{K}(t)+\sum\limits_{n\in\mathcal{N}}\psi^{k}_{n}(t)\leq{U^{K}(t)},\quad{k=1,\cdots,K},\quad\TT,\\
    &\qquad\quad \text{M}_{2}:[{\hat{\psi}}^{l}_1(t),\cdots,{\hat{\psi}}^{l}_{N}(t)]Z^{K}(t)+\sum\limits_{n\in\mathcal{N}}{\hat{\psi}}^{l}_{n}(t)\leq{0},\quad{l=1,\cdots,L},\quad\TT,\\
    &\qquad\quad \text{M}_{3}:Z^{K}(t)\in[0,1],\quad\TT.
    \label{eq:master problem}
    \end{aligned}
\end{equation}
where $\bold{Z}^{K}=\textbf{col}(Z^{K}_{1},\cdots,Z^{K}_{N})$, $\bold{Z}^{K}_{n}=\textbf{col}(Z^{K}_{n}(\tau),\cdots,Z^{K}_{n}(\tau+H))$, and $\bold{U}^{K}=\textbf{col}(U^{K}(\tau),\cdots,U^{K}(\tau+H))$. $\phi_{n}^{k}(t)$ and $\psi_{n}^{k}(t)$ are the dual multipliers of the sub-problem, $\hat{\phi}_{n}^{k}(t)$ and $\hat{\psi}_{n}^{k}(t)$ are the dual multipliers of the feasibility check problem, which are described in in the following. Constraints $\text{M}_{1}$ and $\text{M}_{2}$ are the optimality and feasibility constraints.

\textbf{Feasibility check.}
We can check the feasibility solution of \eqref{eq:master problem} by applying MPC and solving the following optimization problem:

\begin{equation}
    \begin{aligned}
    &F(Z^{K})=\min\limits_{\bold{q},\bold{Y},\bold{Q},\bold{x},\bold{\delta}}\sum\limits_{n\in\mathcal{N}}\sum\limits_{\TT}\delta_{n}(t),\\
    &\text{S.b.}\quad \text{F}_{1}:\sum\limits_{n\in\mathcal{N}}\q=D(t),\quad\TT,\\
    &\quad\qquad \text{F}_{2}:\XP=(1-\Z)(A\X+B\q),\quad{\N},\quad\TT,\\
    &\quad\qquad\text{F}_{3}:\Y\leq{\ZK M}+\delta_{n}(t),\quad\N,\quad\TT:\quad\hat{\gamma}^{1}_{n}(t),\\
    &\quad\qquad\text{F}_{4}:\Y\geq{\X-(1-\ZK)M}-\delta_{n}(t),\quad\N,\quad\TT:\quad\hat{\gamma}^{2}_{n}(t)\\
    &\quad\qquad\text{F}_{5}:\Y\leq{\X+(1-\ZK)M}+\delta_{n}(t),\quad\N,\quad\TT:\quad\hat{\gamma}^{3}_{n}(t),\\
     &\quad\qquad\text{F}_{6}:\Q\leq{\ZK M+\delta_{n}(t)},\quad\N,\quad\TT:\quad\hat{\gamma}^{4}_{n}(t),\\
        &\quad\qquad\text{F}_{7}:\Q\geq{\q-(1-\ZK)M-\delta_{n}(t)},\quad\N,\quad\TT:\quad\hat{\gamma}^{5}_{n}(t),\\
    &\quad\qquad\text{F}_{8}:\Q\leq{\q+(1-\ZK)M+\delta_{n}(t)},\quad\N,\quad\TT:\quad\hat{\gamma}^{6}_{n}(t),\\
        &\quad\qquad\text{F}_{9}:{\q}\leq{(1-\ZK){{\q}^{\max}+\delta_{n}(t)}},\quad\N,\quad\TT:\quad\hat{\gamma}^{7}_{n}(t),\\
        &\quad\qquad\text{F}_{10}:{\q}\geq{(1-\ZK){{\q}^{\min}-\delta_{n}(t)}},\quad\N,\quad\TT:\quad\hat{\gamma}^{8}_{n}(t),\\
    &\quad\qquad\text{F}_{11}:\Q\geq{0},\quad\Y\geq{0},\quad\X\leq{G_{n}},\quad\N,\quad\TT,
    \label{eq:feasibility problem}
    \end{aligned}
\end{equation}
where $\bold{\delta}=\textbf{col}(\delta(\tau),\cdots,\delta(\tau+H))$, and $(\hat{\gamma}^{1}_{n}(t)-\hat{\gamma}^{8}_{n}(t))$ are the dual multipliers of constraints $(\text{F}_{3}-\text{F}_{10})$ of \eqref{eq:feasibility problem} which are obtained by solving the duality problem of \eqref{eq:feasibility problem}. After solving \eqref{eq:feasibility problem}, the first control input at time $\tau$ is applied to the system. Then, the state of the system at $\tau+1$ is obtained. By setting $\tau=\tau+1$, and based on the information of the system at time $\tau+1$, \eqref{eq:feasibility problem} again is solved. This procedure continues until the end of the planning horizon. The optimal solution of \eqref{eq:master problem} is feasible if $F(Z^{K})$ is equal to zero, otherwise it is infeasible. To resolve this issue we must add feasibility cuts to \eqref{eq:master problem}.

\textbf{Feasibility cuts.} In case that the obtained maintenance decisions of \eqref{eq:master problem} are not feasible for the sub-problem, we have to increase $L$ one-step and add the feasibility constraints to \eqref{eq:master problem} using the following equations:
\begin{equation}
\begin{aligned}
\hat{\phi}^{L+1}_{n}(t)=&-\hat{\gamma}^{1}_{n}(t)M+\hat{\gamma}^{2}_{n}(t)M+\hat{\gamma}^{3}_{n}(t)M-\hat{\gamma}^{4}_{n}(t)M\\
&+\hat{\gamma}^{5}_{n}(t)M+\hat{\gamma}^{6}_{n}(t)M+\hat{\gamma}^{7}_{n}(t){q_{n}}^{\max}-\hat{\gamma}^{8}_{n}(t){q_{n}}^{\min},
\label{eq:feasibility cut}
\end{aligned}
\end{equation}

\begin{equation}
\begin{aligned}
\hat{\psi}^{L+1}_{n}(t)=&\hat{\gamma}^{1}_{n}(t)\Y+\hat{\gamma}^{2}_{n}(t)(\X-\Y-M)+\hat{\gamma}^{3}_{n}(t)(\Y-\X-M)\\
&+\hat{\gamma}^{4}_{n}(t)\Q+\hat{\gamma}^{5}_{n}(t)(\q-\Q-M)+\hat{\gamma}^{6}_{n}(t)(\Q-\q-M)\\
&+\hat{\gamma}^{7}_{n}(t)(\q-{q_{n}}^{\max})-\hat{\gamma}^{8}_{n}(t)(-\q+{q_{n}}^{\min}),
\label{eq:feasibility cut1}
\end{aligned}
\end{equation}

\textbf{Sub-problem using the dual decomposition method.}
In case that the maintenance solution of \eqref{eq:master problem} is feasible, the sub-problem is expressed as following quadratic programming problem:

\begin{equation}
\begin{aligned}
    &\max\limits_{\bold{q},\bold{Y},\bold{Q},\bold{x}}\sum\limits_{n\in\mathcal{N}}\sum\limits_{\TT}\big(\mathcal{P}({t})\q-a_{n}({\q})^2-b_{n}\q-\X^{T}\X\big)\\
        &\text{S.b.}\quad\text{S}_{1}:\text{F}_{1},\text{F}_{2}\\
    &\quad\qquad\text{S}_{2}:\Y\leq{\ZK M},\quad\N,\quad\TT,\\
    &\quad\qquad\text{S}_{3}:\Y\geq{\X-(1-\ZK)M},\quad\N,\quad\TT,\\
    &\quad\qquad\text{S}_{4}:\Y\leq{\X+(1-\ZK)M},\quad\N,\quad\TT,\\
     &\quad\qquad\text{S}_{5}:\Q\leq{\ZK M},\quad\N,\quad\TT,\\
        &\quad\qquad\text{S}_{6}:\Q\geq{\q-(1-\ZK)M},\quad\N,\quad\TT\\
    &\quad\qquad\text{S}_{7}:\Q\leq{\q+(1-\ZK)M},\quad\N,\quad\TT,\\
        &\quad\qquad\text{S}_{8}:{\q}\leq{(1-\ZK){{q_{n}}^{\max}}},\quad\N,\quad\TT,\\
        &\quad\qquad\text{S}_{9}:{\q}\geq{(1-\ZK){{q_{n}}^{\min}}},\quad\N,\quad\TT,\\
    &\quad\qquad\text{S}_{10}:\Q\geq{0},\quad\Y\geq{0},\quad\X\leq{G_{n}},\quad\N,\quad\TT.
    \label{eq:subproblem}
    \end{aligned}
\end{equation}
Since Problem \eqref{eq:subproblem} is strongly concave, so we can relax the coupling constraint $\text{F}_{1}$ among the agents using the Lagrangian method and then solve the problem in the dual domain as follows:

\begin{equation}
\begin{aligned}
    &\min\limits_{\bold{\lambda}}\max\limits_{\bold{q},\bold{Y},\bold{Q},\bold{x}}\sum\limits_{\TT}\lambda(t)D(t)+\sum\limits_{n\in\mathcal{N}}\sum\limits_{\TT}\big(\mathcal{P}({t})\q-a_{n}{\q}^2-b_{n}\q-\X^{T}\X-\lambda(t)\q\big)\\
    &\text{S.b.}\quad\text{F}_{2}, \text{S}_{2}-\text{S}_{10}.
    \label{eq:subproblem dual}
    \end{aligned}
\end{equation}
Then, by using the dual decomposition approach, Equation \eqref{eq:subproblem dual} can be solved iteratively. Each agent obtains its optimal decisions and sends them to the central system. The central system updates the dual multiplier using the sub-gradient method and sends it to the agents. These steps continue until the convergence occurs. At iteration $i$ of the dual decomposition algorithm each agent solves the following optimization problem:

\begin{equation}
\begin{aligned}
    &\max\limits_{\bold{q},\bold{Y},\bold{Q},\bold{x}}\sum\limits_{\TT}\big(\mathcal{P}({t})\q-a_{n}({\q})^2-b_{n}\q-\X^{T}\X-\lambda^{i}(t)\q\big)\\
    &\text{S.b.}\quad\XP=(A\X+B\q)-(A\Y+B\Q),\quad\TT\\
    &\qquad\quad\Y\leq{\ZK M},\quad\TT:\quad{\gamma}^{1}_{n}(t),\\
    &\qquad\quad\Y\geq{\X-(1-\ZK)M},\quad\TT:\quad{\gamma}^{2}_{n}(t),\\
    &\qquad\quad\Y\leq{\X+(1-\ZK)M},\quad\TT:\quad{\gamma}^{3}_{n}(t),\\
     &\qquad\quad\Q\leq{\ZK M},\quad\TT:\quad{\gamma}^{4}_{n}(t),\\
        &\qquad\quad\Q\geq{\q-(1-\ZK)M},\quad\TT:\quad{\gamma}^{5}_{n}(t),\\
    &\qquad\quad\Q\leq{\q+(1-\ZK)M},\quad\TT:\quad{\gamma}^{6}_{n}(t),\\
        &\qquad\quad{\q}\leq{(1-\ZK){{q_{n}}^{\max}}},\quad\TT:\quad{\gamma}^{7}_{n}(t),\\
        &\qquad\quad{\q}\geq{(1-\ZK){{q_{n}}^{\min}}},\quad\TT:\quad{\gamma}^{8}_{n}(t),\\
    &\qquad\quad\Q\geq{0},\quad\Y\geq{0},\quad\X\leq{G_{n}},\quad\TT,
    \label{eq:subproblem dual1}
    \end{aligned}
\end{equation}
where $(\gamma_{n}^{1}(t)-\gamma_{n}^{8}(t))$ are the dual multipliers of the constraints of \eqref{eq:subproblem dual1} which are obtained by solving the dual problem of \eqref{eq:subproblem dual1}.

Then, the central system updates the dual multiplier as follows:

\begin{equation}
\lambda^{i+1}(t)=\lambda^{i}(t)+\alpha(t)(\sum\limits_{n\in\mathcal{N}}\q-D(t)).
\end{equation}

After the dual decomposition algorithm converges, the agents implement the obtained control input at iteration time $\tau$ to the system and obtain the system state at $\tau+1$. By setting $\tau=\tau+1$, the agents again solve \eqref{eq:subproblem dual1}. This procedure continues until the end of the planning horizon $\tau=H$. 

\textbf{Optimality cuts.}
If the obtained value of the optimal objective function of \eqref{eq:subproblem dual} is less than the value of the optimal objective function of \eqref{eq:master problem}, this means that we have

\begin{equation}
     [\phi^{K+1}_1(t),\cdots,\phi^{K+1}_{N}(t)]Z^{K}(t)+\sum\limits_{n\in\mathcal{N}}\psi^{K+1}_{n}(t)\leq{U^{K}(t)}+\epsilon,
     \label{eq:terminate}
\end{equation}
where

\begin{equation}
\begin{aligned}
\phi^{K+1}_{n}(t)=&-\gamma^{1}_{n}(t)M+\gamma^{2}_{n}(t)M+\gamma^{3}_{n}(t)M-\gamma^{4}_{n}(t)M\\
&+\gamma^{5}_{n}(t)M+\gamma^{6}_{n}(t)M+\gamma^{7}_{n}(t){q_{n}}^{\max}-\gamma^{8}_{n}(t){q_{n}}^{\min},
\label{eq:feasibility}
\end{aligned}
\end{equation}

\begin{equation}
\begin{aligned}
\psi^{K+1}_{n}(t)=&-\big(\mathcal{P}({t})\q-a_{n}{\q}^2-b_{n}\q-\X^{T}\X-\lambda^{i}(t)\q\big)\\
&{\gamma}^{1}_{n}(t)\Y+{\gamma}^{2}_{n}(t)(\X-\Y-M)+{\gamma}^{3}_{n}(t)(\Y-\X-M)\\
&+{\gamma}^{4}_{n}(t)\Q+{\gamma}^{5}_{n}(t)(\q-\Q-M)+{\gamma}^{6}_{n}(t)(\Q-\q-M)\\
&+{\gamma}^{7}_{n}(t)(\q-{q_{n}}^{\max})-{\gamma}^{8}_{n}(t)(-\q+{q_{n}}^{\min})
\label{eq:feasibility1}
\end{aligned}
\end{equation}

Then, the optimal solution of \eqref{eq:subproblem dual} is the optimal solution of the algorithm. If \eqref{eq:terminate} does not hold, we have to increase $K$ one step and add the optimality cut constraints using \eqref{eq:feasibility} and \eqref{eq:feasibility1} to \eqref{eq:master problem}.

The summary of the proposed method is illustrated in Algorithm \ref{alg:iterative algorithm}.

\begin{algorithm}
	\caption{Distributed algorithm for obtaining the maintenance scheduling and production of manufacturing} 
	\label{alg:iterative algorithm}
	\begin{algorithmic}[1]
		\STATE {\textbf{Input}}: Set $\epsilon$, $K=1$, $L=1$,  $\phi^{K}_{n}(t)$, $\psi^{K}_{n}(t)$, $\hat{\phi}^{L}_{n}(t)$, $\hat{\psi}^{L}_{n}(t)$, $LB(t)=-\infty$, $UB(t)=+\infty$, $\T$, $\N$.\\  
		 \label{alg.l.1}
		 	\STATE Solve \eqref{eq:master problem} and set $LB(t)=U^{K}(t)$, $\T$.
			\STATE Solve \eqref{eq:feasibility problem} using MPC method and obtain $F(Z^{K}(t))$, $\T$.
				\STATE If $F(Z^{K}(t))$, $\T$, is equal to zero go to the next step, otherwise go to Step $9$.
					\STATE Solve \eqref{eq:subproblem dual} using         MPC method and duality approach for $\T$.
					\STATE Update ${\phi}^{K+1}_n(t)$ and ${\psi}^{K+1}_n(t)$ using \eqref{eq:feasibility} and \eqref{eq:feasibility1}.
					\STATE Set $UB(t)=[\phi^{K+1}_1(t),\cdots,\phi^{K+1}_N(t)]Z^K(t) +\sum\limits_{\N}\psi^{K+1}_n(t)$, $\T$.
					\STATE If $UB(t)\leq{LB(t)}+\epsilon$,  $\T$, go to Step $10$, otherwise set $K=K+1$, and go to Step $2$.
					\STATE Update $\hat{\phi}^{L+1}_n(t)$ and $\hat{\psi}^{L+1}_n(t)$ using \eqref{eq:feasibility} and \eqref{eq:feasibility1}, set $L=L+1$, and go to Step $2$.
					\STATE End.
	\end{algorithmic}
\end{algorithm}

\section{Simulation Results}
\label{sec:simulation}

In this section, we implement the proposed algorithm to a manufacturing system with $7$ units. We assume that the price per produced unit is influenced by the demand at the market. When the demand is high, the price is increasing and vise versa. This is a realistic scenario for different types of manufacturing systems, including chemical production plants and other manufacturing systems whose products are traded at markets, such as for example isocyanates that are the basic ingredient of plastics including rigid and flexible foams, elastomers and binders. Furthermore, we assume that the degradation state of the units is evolving slowly so they do not need to perform maintenance very frequently. In other words, we consider small values of the deterioration coefficients in the system dynamic \eqref{eq:dynamic deteroration}. The parameters of the production units are shown in Table \ref{tab:agent model}. 

\begin{table}
  \caption{Unit's model parameters}\centering
 \begin{tabular}{c |c c c c c c c c} 
 & & &  &Unit & & & &  \\
  & $1$ & $2$ & $3$ & $4$  & $5$ & $6$ & $7$ \\
 \hline
 Production cost's parameter ($a$) & 0.0105&0.0090&0.0090&0.0095&0.0085& 0.0075
    &0.0090 \\
 \hline
 Production cost's parameter ($b$) & 15.37 &11.29&8.80&8.00& 11.40&10.45&10.03\\
 \hline
  Maximum capacity $q^{\max}$ &2080&1292&1450&1304&1016&1374
        &1160 \\ 
 \hline
 Minimum capacity $q^{\min}$ & 150&155&175&220&225&130&300\\
 \hline
 Dynamic model parameter ($A$) &1&1&1&1&1&1&1\\
 \hline
 Dynamic model parameter ($B$) & 0.0054& 0.0031& 0.0045& 0.0036&0.0090&  0.0040& 0.0045\\
\end{tabular}
  \label{tab:agent model}
\end{table}

We assume that the contractual product delivery commitments need to be fulfilled at each time. 

\begin{figure}[H]
	\centering
	\includegraphics[width=5in]{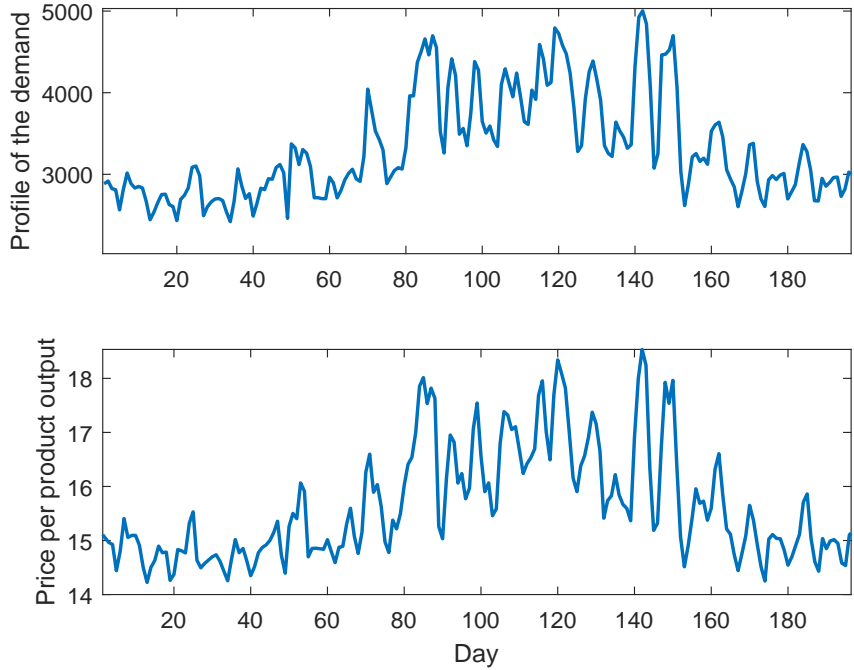}
	\caption{The profile of the demand and the price per product output during $196$ days.}
	\label{demand}
\end{figure}

The results of the joint maintenance and production scheduling are obtained for a period of $196$ days. The demand and the price per production are shown in Figure \ref{demand}.

\begin{figure}[H]
	\centering
	\includegraphics[width=4.5in]{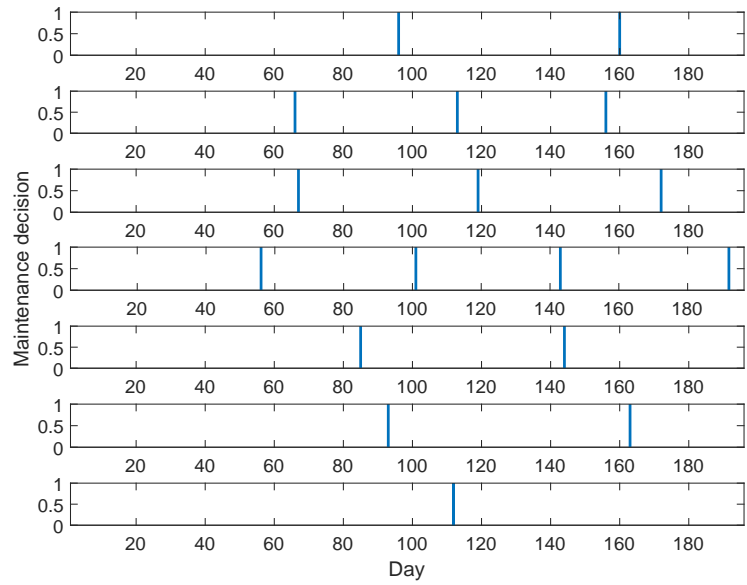}
	\caption{Maintenance scheduling of all units during $196$ days.}
	\label{maintenance}
\end{figure}

\begin{figure}[H]
	\centering
	\includegraphics[width=4.5in]{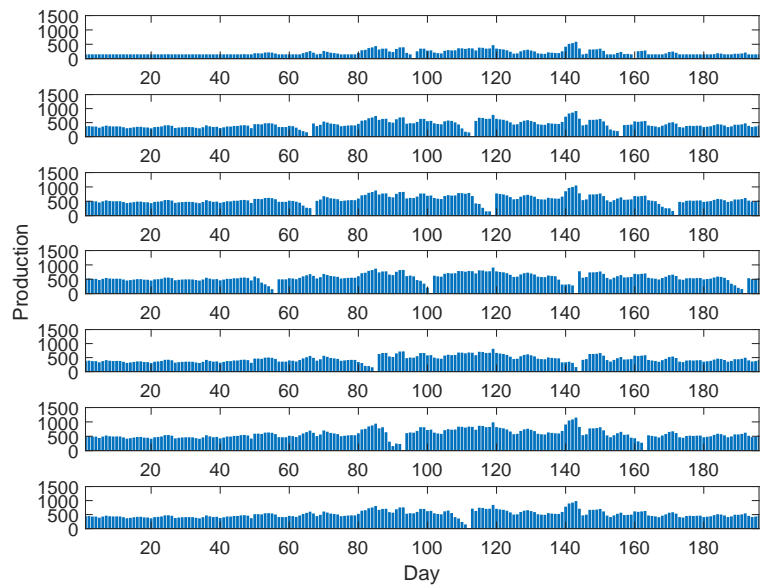}
	\caption{The amount of production of all units during $196$ days.}
	\label{production}
\end{figure}

\begin{figure}[H]
	\centering
	\includegraphics[width=4.5in]{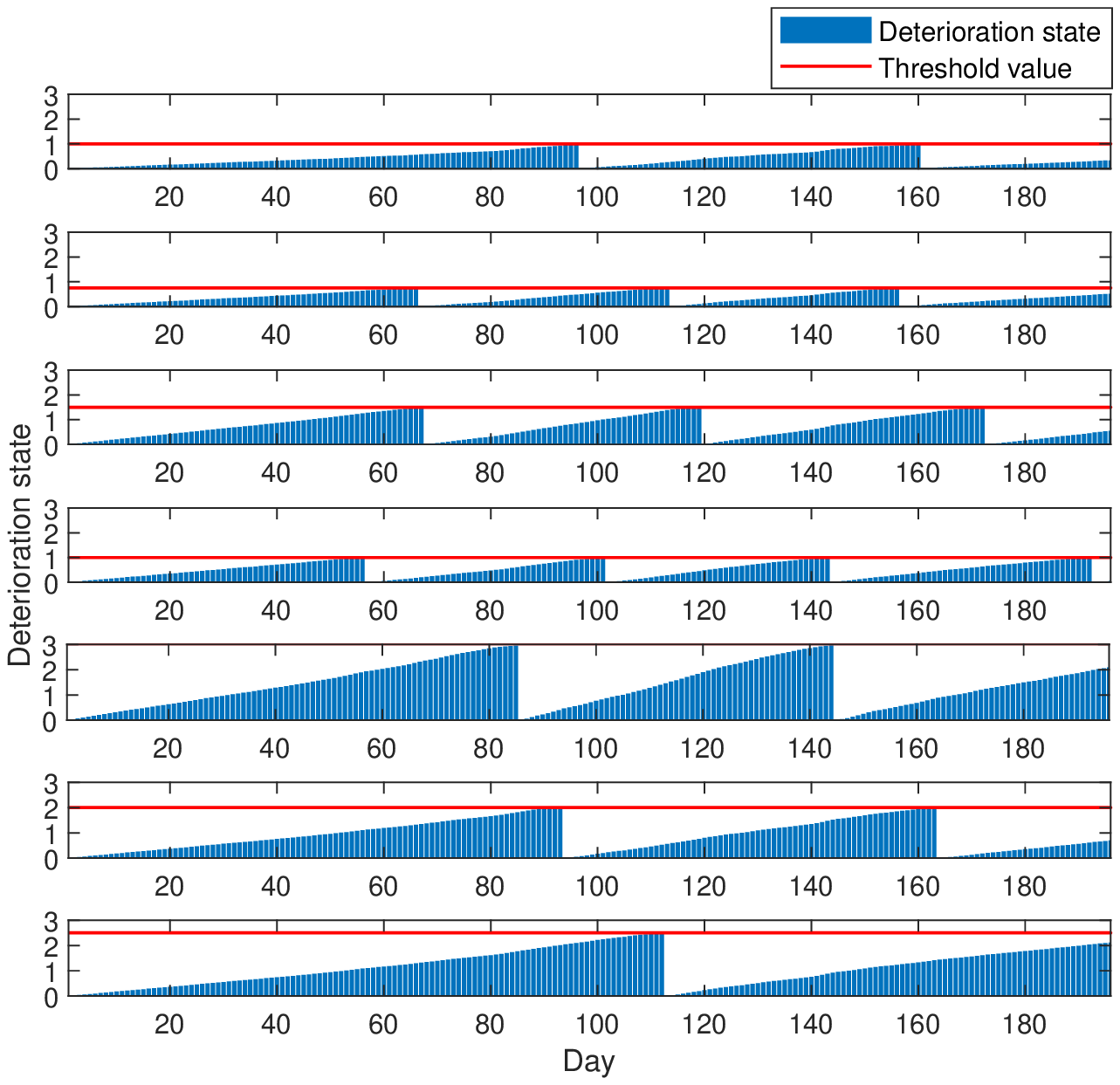}
	\caption{Deterioration state of all units during $196$ days.}
	\label{state}
\end{figure}

The maintenance schedules, production output, and the degradation state of all units are shown in Figure \ref{maintenance}, \ref{production}, and \ref{state}, respectively. As we can see, when the degradation state reaches its threshold value, the units need to perform maintenance. After the maintenance, the state of the system returns to its initial condition. The frequency of the maintenance depends on the threshold values for the states of the different units, the amount of the unit's production, and the dynamic system model of the units. For instance, unit $7$ has a rather high threshold value, resulting in the fact that in the considered time period, the maintenance was performed only once. 

\begin{figure}
	\centering
	\includegraphics[width=4.5in]{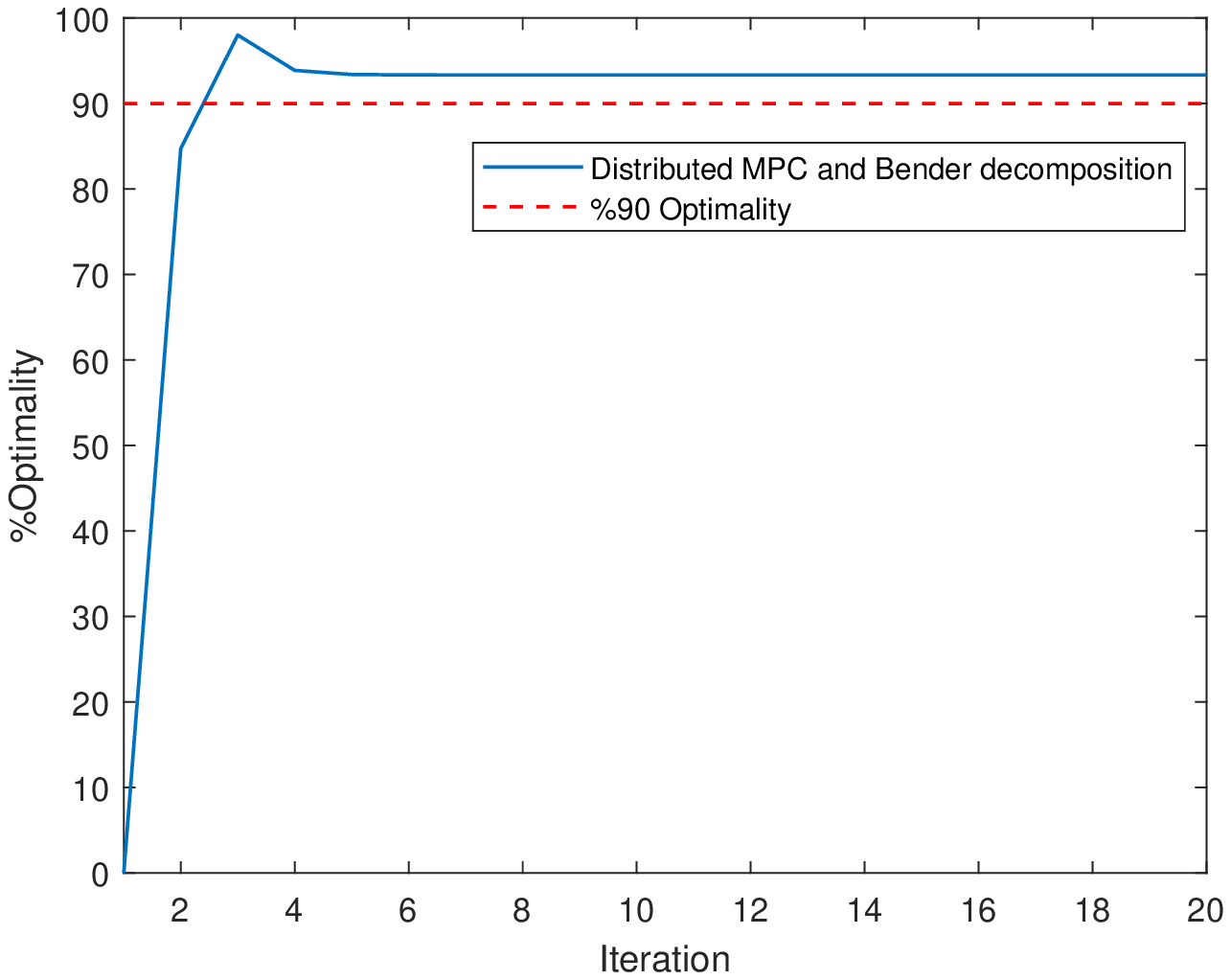}
	\caption{Deterioration state of all units during $196$ days.}
	\label{iteration}
\end{figure}

The convergence of the objective function of the proposed algorithm is shown in Figure \ref{iteration}. As we can see, the objective function converges to its nearly optimal values after about $5$ iterations of the Benders decomposition algorithm.

\begin{figure}
	\centering
	\includegraphics[width=4in]{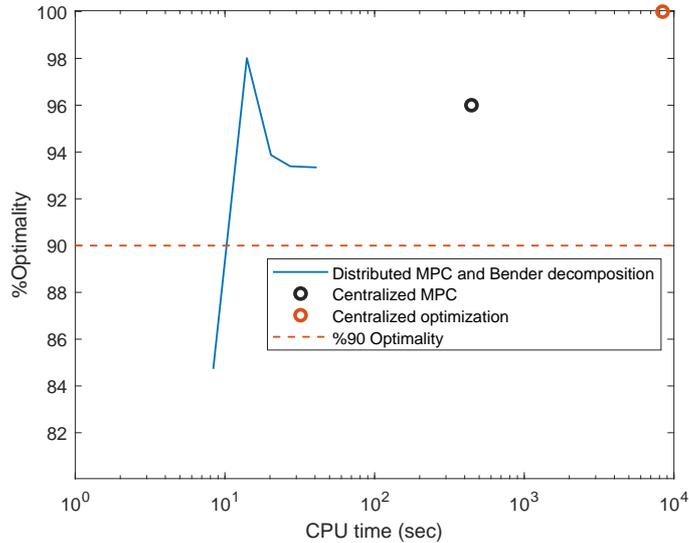}
	\caption{Computation time of the distributed MPC and Benders decomposition algorithm (proposed method), centralized MPC, and the centralized optimization method.}
	\label{cpu}
\end{figure}

Figure \ref{cpu} compares the computational time of the proposed algorithm to the centralized MPC and the centralized optimization method. This figure shows that the computational time of the proposed method is significantly smaller than the one of the other two approaches since each problem has less decision variables so that the solver needs less time to solve them. Also the computational time of the centralized MPC method is less than the centralized optimization since considering a short decision horizon $H$ at each step results in a small-scale optimization which can be solved faster. Moreover, all the methods can achieve more than $90\%$ optimality of the solution. Is worth noting that the optimal solution of the centralized MPC is slightly higher (by 2.66\%) than the proposed algorithm with a much higher computational burden. Hence, by applying the proposed algorithm we make a trade-off between the optimality gap and the computational time. Moreover, in our proposed algorithm, the central system does not need to know all the information of the agents. Thereby, the privacy of the agents is preserved. This feature may be particularly relevant in contexts where agents are independent entities and do not belong to the same company.

\section{Conclusion}
\label{sec:conclusion}

In this paper, we address the joint maintenance and production scheduling for  manufacturing systems with  dynamic degradation states. In order to solve the problem in a distributed way, we propose a novel framework based on the Benders decomposition and the distributed MPC applying the dual decomposition approach. The simulation results show that the solution of the algorithm is close to the optimal value and is computationally more efficient compared to the distributed MPC and the centralized optimization method.

As future research direction, the proposed framework can be extended by integrating the uncertainty in the dynamics of the system. This will require stochastic optimization approaches. Moreover, modeling the demand uncertainty can be considered as another future research direction.
\section*{Acknowledgment}
The contribution of Olga Fink was funded by the Swiss National Science Foundation (SNSF) Grant no. PP00P2\_176878. 

\bibliographystyle{agsm}
 \bibliography{bib_items}
\end{document}